# Smart Grid Architecture with High Proportion of Energies Utilization

*Javeria Noor 罗家伟 (214618005)*

*School of Automation, Department of Control Science and Engineering, Central South University, Changsha China*

I. **Abstract:**

DC power generation is an emerging trend and has been preferred due to its low cost and low in system power losses within local distribution grid. The theme of this paper is the indigenous design of a DC standalone micro grid which will stabilize the fluctuating power generated from hybrid energy resources and manage its power distribution. The power is distributed through a DC distribution line and converted to the required AC or DC form by converters placed near loads. This project includes power generation, distribution and management strategies for a sustainable micro grid primarily powered by wind and solar energy. DC transmittable power can increase the system efficiency up to 10% as compared to AC.

II. **Keywords:**

III. **Introduction:**

Electricity has delivered numerous things that unquestionably have made numerous wonders and life would appear to be so intense without it. Electricity makes our light, electronic gadgets, for example, PCs and TV, and a large group of basic administrations that we underestimate. The interest of power is expanding quickly regardless of its interfered with supply. The explanation is that there is a sharp increment in populace Renewable energies assets incorporate breeze, sun powered, biomass and geothermal energy sources. This implies all energy asset that reestablish themselves inside a brief timeframe or are for all time accessible. In the course of the most recent thirty years, Asia has become a significant player on the worldwide scene. In view of these improvements, power request is relied upon to increment 8% consistently until 2015. As the world awakens to the truth of environmental change, power will progressively need to originate from sustainable sources, for example, wind and sunlight based. Pakistan is in a decent situation to abuse these because it has inexhaustible wind and sun. Availability of energy in any country has a strong relationship with its economic and social stability. Pakistan, despite the enormous potential of its energy resources, remains energy deficient because the use of renewable energy resources is not common yet. According to the Authors of [1, 2] hybrid storage and Energy production technique increases system efficiency and lowers battery expense Pakistan is confronting a power emergency. Other than the low producing limit, the absence of upkeep of open claimed power plants and underinvestment in the energy sector, halfway actualized strategies together with basic changes are one of the most intense components of the Energy crisis. Pakistan's energy production can't satisfy present and future needs. Urban regions of Pakistan have been confronting a phenomenal

Power crisis since last numerous years. Other than the expanding requests of urban territories, a huge piece of the country populace doesn't have the Power station since they are either excessively remote, or it is seen as too costly to even consider connecting their towns to the national grid station. Thus, the goal of Electrifying rural areas of country won't be achieved by expanding the issue national grid network, with its old, brought together generators, its transmission losses and its unpaid bills and revealed obligation. The modernized, 21st century Pakistan should be less dependent on long Transmission Lines. For supplying electricity to remote areas, the method of using DC microgrid is proposed. Micro-Grid is a homegrown energy network that utilize renewable energy sources and storage systems. This can be cascaded with national grid or can work independently when there is a power failure at the main station, and provide the uninterrupted supply. [3, 4] Pakistan is already leading in small-scale hydro-electric systems. but not all villages have access to hydropower. To electrify many of its villages, Pakistan needs to accelerate its application of Renewable Energy Resources to supply whole villages or clusters of villages by developing the micro-grids. Pakistan has high potential of renewable energy sources. So, in this regard a solution is proposed, "To electrify the rural areas an efficient system can be designed to utilize the hybrid and distributive renewable energy resources using the advanced techniques of Power Electronics".

## IV.   Literature Review:

Microgrids have a long history. Truth be told, Thomas Edison's first power plant built in 1882 – the Manhattan Pearl Street Station – was basically a microgrid since our unified framework was not yet settled. By 1886, Edison's firm had introduced 58 Direct Current (DC) microgrids.

The basic idea of a "microgrid" can be summarized along these lines: a coordinated energy framework comprising of conveyed energy assets and different electrical burdens working as a solitary, independent network either in corresponding to or "islanded" from the current utility power lattice. In the most well-known design, circulated energy assets are integrated on their own feeder, which is then connected to the framework at a solitary purpose in like manner coupling. Microgrids can be seen as the structure squares of the smart grids, or as an elective way as the much-advertised brilliant "Super Grid." As with most microgrids, Edison's electric framework was little with confined age and a restricted conveyance organization [5]. It just served a couple of squares toward any path due to the limitations of the DC transmission organization, which accommodates the present meaning of a microgrid. Strangely, it likewise included batteries to give energy stockpiling. Microgrids with energy stockpiling is an exceptionally intriguing issue today, so by and by Edison was thinking ahead. His framework additionally provided heat from its steam generation to structures around its buildings. Today we call this consolidated heat and power - CHP. By and by Edison's thoughts were right on the money. CHP has demonstrated to be truly important.

In June 2015, an understudy from Mazandaran University, Mazandaran, Iran, present the advancement of instructive DC microgrid stage which incorporates well known sustainable power sources and hybrid storage systems [6-8]. This lab-scale stage gave an instructive climate to senior understudies and graduate understudies to participate in research facility tests and to comprehend and foster novel thoughts for DC power framework applications.
Energy has been the life blood for proceeding with progress of human civilization. Since the start of modern upset around two centuries prior, the worldwide energy utilization has expanded huge amounts at a time to speed up the human expectation for everyday comforts on this planet, especially in the industrialized countries. Per-capita energy utilization, especially in electrical structure, has been an indicator of a country's monetary success. Today, the significant measure of power is produced by petroleum derivative and thermal energy stations [9]. Non-renewable energy source plants make ecological contamination issue while the atomic plants have wellbeing issue. These contamination and security issues are currently causing main pressing issues in our general public. How might we persistently work on our expectation for everyday comforts, and simultaneously, live in a cleaner and more secure climate? Environmentally friendly power sources, and potentially the combination energy in future, can somewhat relieve these issues, yet protection of energy with more productive utilization of power is an unmistakable technique for tackling this issue. Power gadgets and drives innovation is as of late appearance a significant effect in energy protection other than its overall job in worldwide modern mechanization. The role of power electronics and drives technology, which has gone through rapid advancement during the last four decades will be reviewed in a broader perspective [10]. The recent advances of power semiconductor devices, converter topologies, variable frequency drives, control and estimation techniques will be discussed, and technology trends will be outlined wherever possible [11].

A Power Electronic Circuit consists of a power circuit that converts the input power (according to the load requirement) and delivers to the load while the electronic circuit is a Controller Circuit that compares the output with reference and generates a resulting PWM signal accordingly, to control the power switches.

In Power electronic circuits, where efficiency is not main concern, resistors, transistors (in linear mode) and other non-magnetic elements are used. But in Power Converters, efficiency is the major concern so transistors (in switch mode), inductors and transformers are used. Resisters are not used due to power losses that decreases the efficiency. Types of converters are:

| | | |
|---|---|---|
| DC-DC | Buck & Boost Converters | Only magnitude of output can be controlled. |
| AC-DC | Rectifiers Following lowpass Filters | Only magnitude of output can be controlled. |
| DC-AC | Power Inverters (H-bridge Topology) | Both magnitude and freq. can be controlled. |
| AC-AC | Step up & down Transformers | Both magnitude and freq. can be controlled. |

**Table 1: Types of Converters**

V. **System Configuration:**

The main idea of this research is to design and develop a prototype for a localized DC-Microgrid with Hybrid Distributive Power Generation System. The power extracted from environment through renewable energy resources is always in the raw form, this power requires some sophisticated power management and control techniques to convert this power into useable form [12-14]. Moreover, the output of the most of renewable energy resources is the DC Voltage so the strategy behind this research is to use this extracted Power in DC form rather than converting it into AC. After stabilizing the DC Power to some standard Voltage and current, it is distributed locally over a distribution line called DC bus. So, the control and management of this research a consist of DC-DC Converters and Feedback Controllers which regulate and maintain the Voltage to a standard value of DC Bus. As per the technicalities of the research, it is divided into different modules[15]. These modules are designed and tested individually and then integrated to result into a prototype of DC Microgrid with hybrid distributive renewable energy resources. Block Scheme of proposed system is shown in Figure 1.

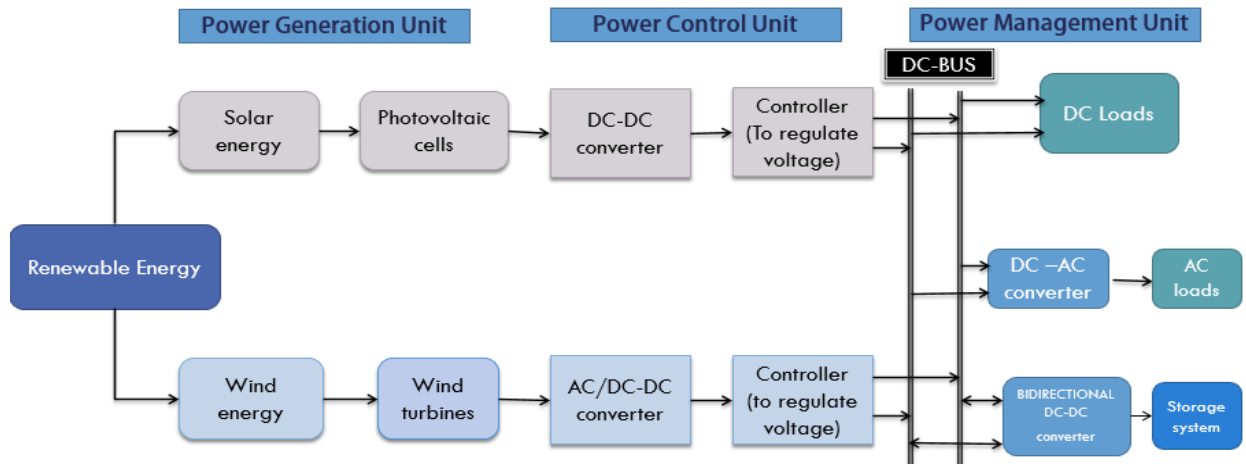

**Figure 1: Block Diagram**

### A. Power Generation Unit

Proposed system in this paper has localized power generation unit which means energy will be generated at the location of utilization. Transmission and distribution losses in the system will be reduced power is generated and distributed near loads. The source of Power generation is Renewable energy. Out of so many renewable energy resources available, we have designed a prototype for the two renewable energy resources that are "Wind Energy and Solar Energy" because they are available everywhere. For Solar energy, Solar panels are being used and tested with the designed Dc-Dc Converters and Controllers. For Wind Energy, a wind turbine simulator is being designed and tested with its respective circuitry.

**I.    Wind Module:**

The simulator that is used for the windmills is coupled DC motor generator assembly, which exactly fulfill the properties and the output of the windmills with respect to variable speed and intensity of wind or air. A motor generator (M-G) assembly refers to a merged device consisting of a motor and a generator mechanically coupled through the common shaft. Practically a motor generator set is a system where a motor and a generator are cascaded in a single circuit. This device is used to convert electrical as well as mechanical power from one form to another [16].

**II.    Solar Module**

Sun powered energy source is seemingly the cleanest, most dependable type of sustainable power source accessible, and it very well utilized to help power your home or business. Sun controlled photovoltaic (PV) boards convert the sun's beams into power by energizing electrons in silicon cells utilizing the photons of light from the sun. This power would then be utilized as power source for your home or business [17]. Maximum power point technique (MPPT) is applied to get greatest force out of the PV module

There are basically two MPPT tracking Methods
- Direct Method
- Indirect Method

Direct Method involves
- Perturb and observed method
- Incremental conduction method

Indirect Method involves
- Fixed Voltage Method
- Fractional open circuit voltage method

MPPT Algorithm shown in Figure 2

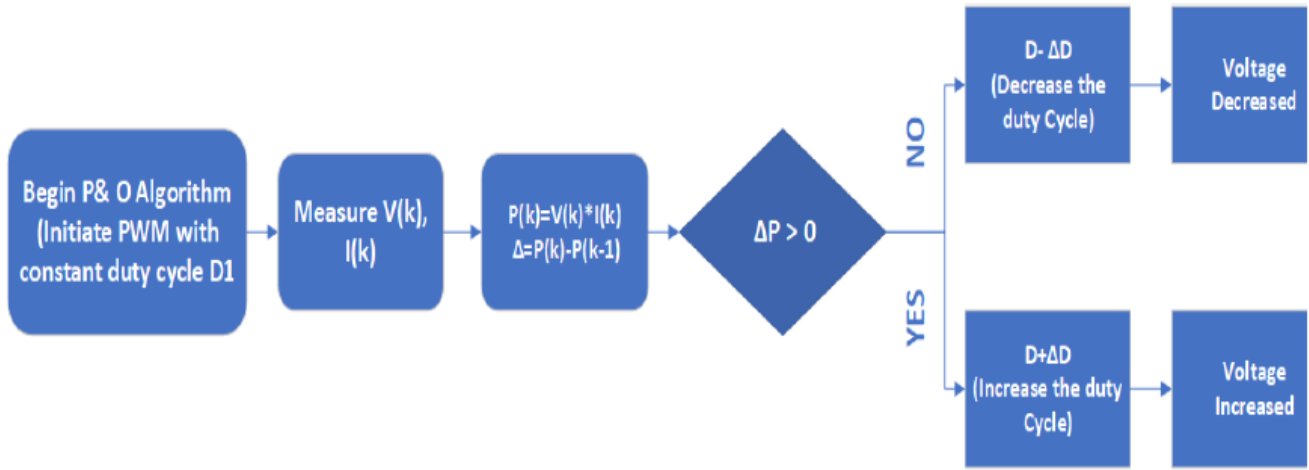

**Figure 2: MPPT Algorithms**

### MPPT Algorithm
In the proposed method in this research, MPPT Algorithm is developed as
- PWM of constant duty cycle D is initiated, measure voltage and current at this duty cycle D. using this voltage and current power is calculated and after each 2 sec delay powers is measured again, then measure the difference of present and previous power
- If Difference of power is greater than zero, then perturbation is added in duty cycle and voltage increased further.
- If Difference of power is less than zero, then added perturbation in duty cycle is subtracted and voltage is decreased, and the previous point is maximum power point.

### B. Power Control Unit
The fluctuating Power extracted from Solar panels and Wind turbines is then regulated and controlled through DC-DC Converters. The Converters are designed and developed in such a way that they efficiently step up or step down the DC voltage as per the value of Standard DC bus Voltage. Energy generated from all the sources must be stable and continues even when the loads are varied [18-20]. To tackle with the sources side variations in voltage and the load side variations, DC-DC Converters are controlled through feedback Controller. The Analog Proportional Integral (PI) Controllers are being used with DC-DC Converters. The Controller is designed and derived from the transfer functions of DC-DC Converters using root locus techniques. While the transfer function of DC-DC Converters is derived through the techniques of small signal AC modeling[21].
On the chance if mathematical model of the plant can be inferred, at that point it is conceivable to apply different procedures for deciding parameters of the controller that will meet the transient and consistent state details of the close loop framework. In any case, if the numerical model can't be effectively gotten, at that point an expository or computational way can be used to deal with the parameters of a PI controller [22].

Ziegler and Nichols [23] suggested rules for tuning PID controllers (meaning to set values and) based on experimental step responses or based on the value of that results in marginal stability when only proportional control action is used. These methods are useful when mathematical models of plants are not known. Such rules suggest a set of values of and that will give a stable operation of the system Transfer function of the Control System is shown by Equation 1

$$G_c(s) = K_p \left(1 + \frac{1}{T_i s} + T_d s\right) \qquad (1)$$

### C. Power Management Unit

Another major feature of this research is the Power management unit. It consists of storage system with bi-directional Converter. The functionality of this unit is to manage the power flow in such a way that when there is excess energy generating from sources the power will flow from DC-Bus to Storage system through Bidirectional Converter to store the surplus energy. And when there comes an increase in load demand and DC-Bus in insufficient to supply enough energy, the stored power will then flow from storage system to the DC-Bus. The charging and discharging of storage system are controlled through current and voltage sensors being used with a digital Controller.

#### 1. Local Distribution Lines

DC Bus is the Distribution line that connects the power sources to loads through different Dc-Dc Converters. In today's electrical grid, high voltage DC (HVDC) [24] transmission lines are one of the most efficient ways of transmitting substantial amounts of power over long distances. HVDC transmission lines are suitable for both overhead lines, underground lines and subsea lines. The use of HVDC in transmission lines proves to be much more efficient compared to HVAC transmission lines, in terms of economy and environment [25]. The DC distribution lines can be two wire or three wire as per the standards. In our prototype, we have used two wire DC transmission system.

#### 2. DC-DC Converters

A dynamic analysis is performed to select a power conversion unit - dc-dc converter - which can provide energy with high quality standards from the wind turbine and solar panel to the DC microgrid. The DC/DC converter that is used in wind module to manage and control the output of windmill [20, 26, 27].

#### 3. Buck Converter

A DC-to-DC power converter that steps down voltage (while stepping up current) from its input (supply) to its output (load). It is containing at least two semiconductors, modern buck converters frequently replace the diode with a second transistor used for synchronous rectification and at least one energy storage element, a capacitor, inductor, or the two in combination. Filters made of capacitors (sometimes in combination with inductors) are normally added to such a converter's output (load-side filter) and input (supply-side filter) to reduce voltage ripple. Schematic of Buck converter is shown in Figure 3

The buck converter is designed to find the desired output and find the values of R L and C using small signal AC modeling to obtain the transfer functions which is as under Equation 2

$$G_{rd}(s) = \frac{v}{D} \left[\frac{1}{1+s(4/R)+s^2(1/LC)}\right] \qquad (2)$$

Matlab Simulation of transfer function of closed loop and open loop is as :

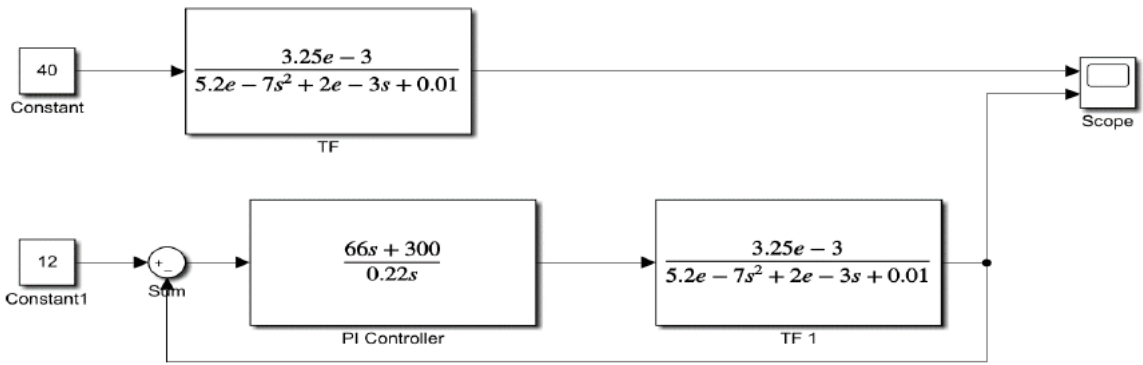

Figure 3: Open Loop and Closed Loop Transfer Function of Buck Converter

To control this buck converter, a PI controller is designed by calculating the $K_P$ and Ki from transfer function of converter using root locus technique. Transfer function is shown by Equation 3

| S2 | $5.2 \times 10^{-7}$ | 0.01 |
|---|---|---|
| S1 | $2 \times 10^{-3}$ | $k_p$ |
| S0 | $\dfrac{(2 \times 10^3)(0.01) - 5.2 \times 10^{-1} k_p}{2 \times 10 - 3}$ | 0 |

So $\quad\quad\quad\quad\quad\quad K_p = 15 \quad\quad Ki = 0.002$

$$G_c(S) = \frac{0.03S + 15}{0.002S} \quad\quad (3)$$

Output waveform is shown in Figure 4

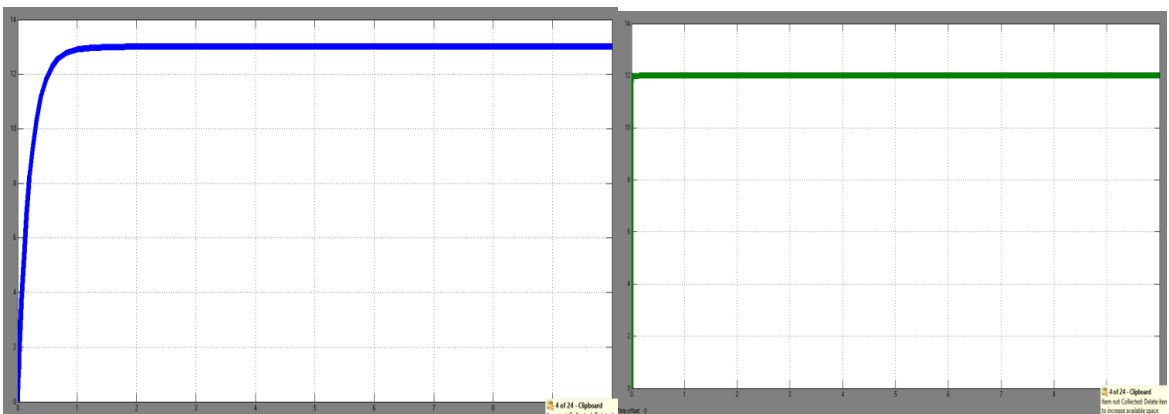

Figure 4: Output Response of Closed Loop and open loop Transfer Function of Buck Converter

### 4. Boost Converter

A DC-to-DC power converter [28] that steps up voltage (while stepping down current) from its input (supply) to its output (load). It is a type of switched-mode power supply [29] (SMPS) containing two semiconductors (a diode and a transistor) and at least one energy storage element: a capacitor, inductor, or the two in combination. Schematic of Boost converter along with its output is shown in Figure 5-6. The boost converter is designed to find the desired output and find the values of R L and C using small signal AC modeling to obtain the transfer functions which is shown in Equation 4.

$$G_{rd}(s) = \frac{D'^2 - s(L/R)}{D'^3 + s(L/R)D' + s^2(LD'C)} \quad (4)$$

To control this boost converter, a PI controller is designed by calculating the Kp and Ki from transfer function of converter using root locus technique.

| | | |
|---|---|---|
| S2 | $5.2 \times 10^{-7}$ | 0.01 |
| S1 | $2 \times 10^{-3}$ | $k_p$ |
| S0 | $\dfrac{(2 \times 10^3)(0.01) - 5.2 \times 10^{-1} k_p}{2 \times 10 - 3}$ | 0 |

So

$$K_P = 0.02715 \qquad Ki = 0.004$$

Transfer function is shown by Equation 5.

$$G_c(s) = \frac{10.8 \times 10^{-5} S + 1.81 \times 10^{-2}}{0.004 S} \quad (5)$$

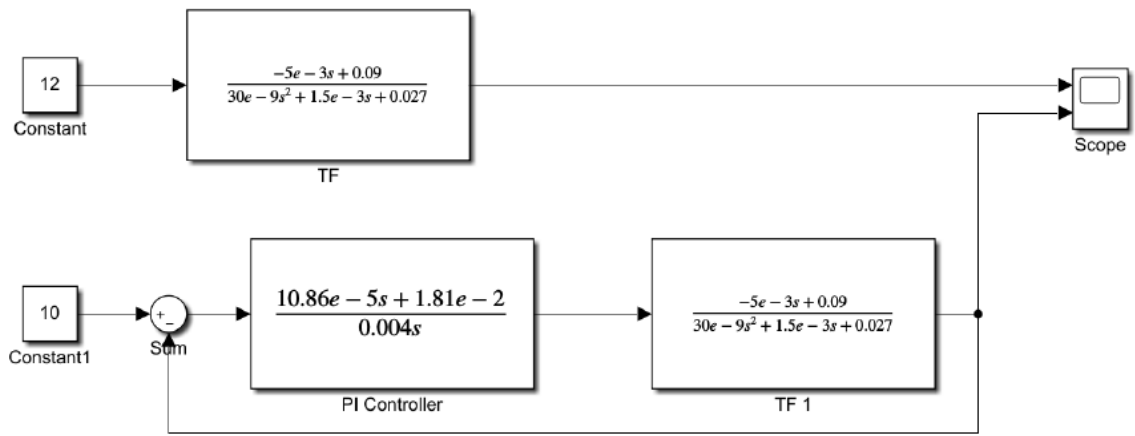

**Figure 5: Open loop and Closed Loop Transfer Function of Boost Converter**

The Output Waveform is shown in Figure 6

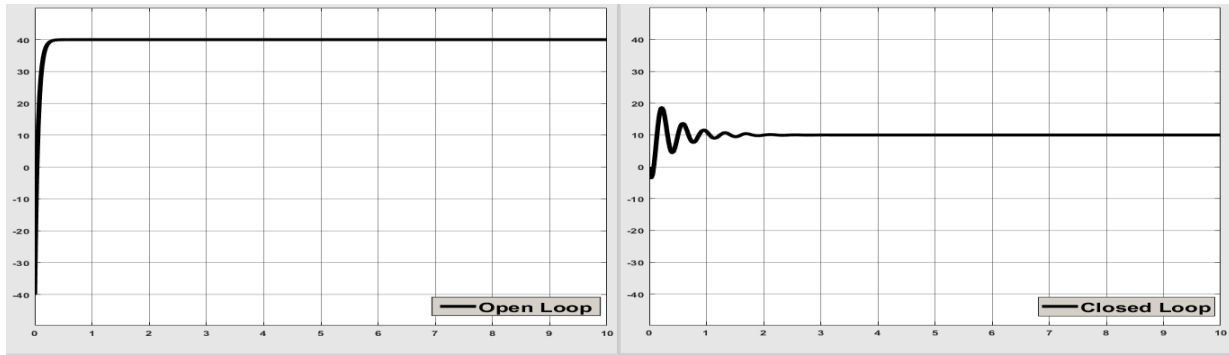

**Figure 6: Output Response of Open loop and closed loop Transfer Function of Boost Converter**

5. **Bi-Directional Converter**

Bidirectional dc-dc converters (BDC) [30-32]have as of late got a great deal of consideration because of the expanding need to frameworks with the capacity of bidirectional energy movement between two dc buses. Aside from customary application in dc engine drives, new utilizations of BDC remember energy stocking for sustainable power source frameworks, power device vitality frameworks, hybrid electric vehicles (HEV) and uninterruptible power supplies (UPS). The fluctuating nature of power source assets, like wind and sun, makes them unacceptable for independent activity as the sole source of energy. A typical answer for this issue is to utilize an energy storage gadget other than the sustainable power source asset to make up for these changes and keep up a smooth and lossless energy transmission to load [33, 34].

The bidirectional DC-DC converter acts a buck converter one way and as a Boost converter the other way. Power MOSFETS are utilized as a switching device in the circuit. The activity of the converter is constrained by checking the DC bus voltage and the voltage of the battery. The primary motivation behind the bidirectional DC-DC converter is to keep up the voltage of the DC bus moderately steady at a reference value.

`The DC-DC converter working modes can be separated into three modes:

**Mode 1:**

The DC-DC converter acts in buck mode when the DC transport voltage is more than the set reference. DC-DC converter controls the current to charge the battery in this mode.

**Mode 2:**

The DC-DC converter acts in support mode when the DC transport voltage falls underneath the set reference. Battery is discharged in this mode.

**Mode 3:**

At the point when the battery is completely energized, the DC-DC converter closes to abstain from harming the battery. Hardware for bi-directional convertor is appear in Figure 7

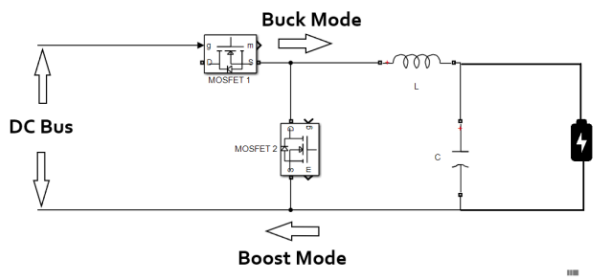

**Figure 7: Bidirectional Converter**

### D. Final Integration and Results

In the final phase Parallel association of power converters are made based on Output Power and Output Voltages. This offers the chance to utilize them in an "independent" design or associated in equal or in arrangement

In a perfect world, the modules ought to be associated in parallel configuration as appeared in Figure 8. In the design the complete yielded power is shared by the two converters. Since the force may not be equitably shared between the two converters, it is prescribed to utilize converters that have greatest evaluated yield power of around 75% of the maximum power. By this, load on either converter won't surpass its individual appraised most extreme rating. Ultra-quick diode of 30A current rating is utilized to adapt up to high current evaluations, a diode should be associated in arrangement with the positive yield of each resembled module. This will forestall a unit that has short disappointment at the Output from shorting out the whole burden

- As two renewable energy resources are used in our research i.e., Wind and Solar. The outputs from both modules are integrated to the DC-Bus together using parallel connection explained above.
- The Storage module is also connected to the DC bus in parallel with the loads
- The proposed DC microgrid system is stable and efficient and can be used in remote areas where National grids are not present or either these areas are far away from the grids.

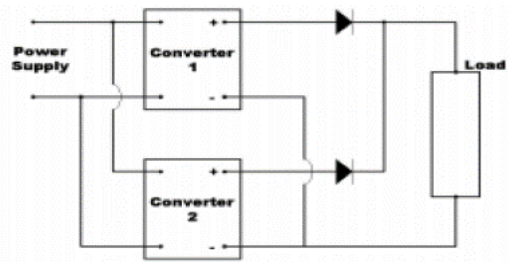

## VI. Future Model:

For future work, Following are the changes that can be implemented in present work.

I. Digital Controller can be used instead of analog controllers to acquire more efficient results.
II. Renewable Energy Resources other than Wind and Solar can be integrated in this system.
III. In storage module, storage capacitors may be used instead of Batteries.
IV. The source of Auxiliary Power supplies can be DC-Bus Voltage.
V. AC-Loads can also be tested with this system by placing an inverter after DC-Bus.
VI. This System can be tied with AC grid through common coupling point.
VII. Buck-Boost Converters can be used instead of simple Boost or Buck Converters to tackle the variations efficiently.

## VII. Conclusion:

This Research results a DC Microgrid with Hybrid Distributive Renewable Energy Resources. The energy from the two resources i-e- Wind and Solar is mixed through advanced Power Electronics techniques. DC-DC Converters are used to stabilize the output voltages of renewable energy resources to the standard DC-Bus Voltage. The Outputs of the DC-DC Converters are controlled through analog PI Controller. It has been proved through various research that DC Power Generation and Distribution System is 10% more efficient than AC System.

**References:**

[1] G. M. Masters, "Wind power systems," 2004.

[2] R. Sathishkumar, S. K. Kollimalla, and M. K. Mishra, "Dynamic energy management of micro grids using battery super capacitor combined storage," in *2012 Annual IEEE India Conference (INDICON)*, 2012: IEEE, pp. 1078-1083.

[3] G. Chicco and P. Mancarella, "Distributed multi-generation: A comprehensive view," *Renewable and sustainable energy reviews,* vol. 13, no. 3, pp. 535-551, 2009.

[4] E. Planas, J. Andreu, J. I. Gárate, I. M. De Alegría, and E. Ibarra, "AC and DC technology in microgrids: A review," *Renewable and Sustainable Energy Reviews,* vol. 43, pp. 726-749, 2015.

[5] C. L. Moreira, F. O. Resende, and J. P. Lopes, "Using low voltage microgrids for service restoration," *IEEE Transactions on Power Systems,* vol. 22, no. 1, pp. 395-403, 2007.

[6] M. Glavin, P. K. Chan, S. Armstrong, and W. Hurley, "A stand-alone photovoltaic supercapacitor battery hybrid energy storage system," in *2008 13th International power electronics and motion control conference*, 2008: IEEE, pp. 1688-1695.

[7] W. Li and G. Joos, "A power electronic interface for a battery supercapacitor hybrid energy storage system for wind applications," in *2008 IEEE Power Electronics Specialists Conference*, 2008: IEEE, pp. 1762-1768.

[8] G. Zhang, X. Tang, and Z. Qi, "Research on battery supercapacitor hybrid storage and its application in microgrid," in *2010 Asia-Pacific Power and Energy Engineering Conference*, 2010: IEEE, pp. 1-4.

[9] H.-L. Tsai, C.-S. Tu, and Y.-J. Su, "Development of generalized photovoltaic model using MATLAB/SIMULINK," in *Proceedings of the world congress on Engineering and computer science*, 2008, vol. 2008: San Francisco, USA, pp. 1-6.

[10] Z. Zheng, X. Wang, and Y. Li, "A control method for grid-friendly photovoltaic systems with hybrid energy storage units," in *2011 4th International Conference on Electric Utility Deregulation and Restructuring and Power Technologies (DRPT)*, 2011: IEEE, pp. 1437-1440.

[11] K. Yoshimoto, T. Nanahara, and G. Koshimizu, "Analysis of data obtained in demonstration test about battery energy storage system to mitigate output fluctuation of wind farm," in *2009 CIGRE/IEEE PES Joint Symposium Integration of Wide-Scale Renewable Resources Into the Power Delivery System*, 2009: IEEE, pp. 1-1.

[12] R. Hebner, J. Beno, and A. Walls, "Flywheel batteries come around again," *IEEE spectrum,* vol. 39, no. 4, pp. 46-51, 2002.

[13] R. Yinger, "Behavior of Capstone and Honeywell Microturbines During Load Changes," ed.

[14] A. Engler, "Applicability of droops in low voltage grids," *International Journal of Distributed Energy Resources,* vol. 1, no. 1, pp. 1-6, 2005.

[15] N. Jayawarna, N. Jenkins, M. Barnes, M. Lorentzou, S. Papthanassiou, and N. Hatziagyriou, "Safety analysis of a microgrid," in *2005 International Conference on Future Power Systems*, 2005: IEEE, pp. 7 pp.-7.

[16] S. Papathanassiou, D. Georgakis, N. Hatziargyriou, A. Engler, and C. Hardt, "Operation of a prototype Micro-grid system based on micro-sources equipped with fast-acting power electronics interfaces," *Available On line http://microgrids. power. ece. ntua. gr,* 2004.

[17] C. Andrews, F. Arsanjani, M. Lanier, J. Miller, T. Volkmann, and J. Wrubel, "Special considerations in power system restoration," *IEEE Trans. Power Syst,* vol. 7, no. 4, pp. 1419-1427, 1992.

[18] Y. W. Li and C.-N. Kao, "An accurate power control strategy for power-electronics-interfaced distributed generation units operating in a low-voltage multibus microgrid," *IEEE Transactions on Power Electronics,* vol. 24, no. 12, pp. 2977-2988, 2009.

[19] N. Eghtedarpour and E. Farjah, "Power control and management in a hybrid AC/DC microgrid," *IEEE transactions on smart grid,* vol. 5, no. 3, pp. 1494-1505, 2014.

[20] Y. Ito, Y. Zhongqing, and H. Akagi, "DC microgrid based distribution power generation system," in *The 4th International Power Electronics and Motion Control Conference, 2004. IPEMC 2004.*, 2004, vol. 3: IEEE, pp. 1740-1745.

[21] Y.-P. Hsieh, J.-F. Chen, T.-J. Liang, and L.-S. Yang, "A novel high step-up DC–DC converter for a microgrid system," *IEEE Transactions on Power Electronics,* vol. 26, no. 4, pp. 1127-1136, 2010.

[22] R. K. Mudi and N. R. Pal, "A self-tuning fuzzy PI controller," *Fuzzy sets and systems,* vol. 115, no. 2, pp. 327-338, 2000.

[23] B. R. Copeland, "The design of PID controllers using Ziegler Nichols tuning," *Internet: http://educypedia. karadimov. info/library/Ziegler_Nichols. pdf,* 2008.

[24] N. Flourentzou, V. G. Agelidis, and G. D. Demetriades, "VSC-based HVDC power transmission systems: An overview," *IEEE Transactions on power electronics,* vol. 24, no. 3, pp. 592-602, 2009.

[25] R. Rudervall, J. Charpentier, and R. Sharma, "High voltage direct current (HVDC) transmission systems technology review paper," *Energy week,* vol. 2000, pp. 1-19, 2000.

[26] H. Lotfi and A. Khodaei, "AC versus DC microgrid planning," *IEEE Transactions on Smart Grid,* vol. 8, no. 1, pp. 296-304, 2015.

[27] S. Beheshtaein, R. M. Cuzner, M. Forouzesh, M. Savaghebi, and J. M. Guerrero, "DC microgrid protection: A comprehensive review," *IEEE Journal of Emerging and Selected Topics in Power Electronics,* 2019.

[28] J. C. Rosas-Caro, J. M. Ramirez, F. Z. Peng, and A. Valderrabano, "A DC–DC multilevel boost converter," *IET Power Electronics,* vol. 3, no. 1, pp. 129-137, 2010.

[29] S. Singh, B. Singh, G. Bhuvaneswari, and V. Bist, "Power factor corrected zeta converter based improved power quality switched mode power supply," *IEEE Transactions on industrial electronics,* vol. 62, no. 9, pp. 5422-5433, 2015.

[30] H. Tao, A. Kotsopoulos, J. L. Duarte, and M. A. Hendrix, "Family of multiport bidirectional DC–DC converters," *IEE Proceedings-Electric Power Applications,* vol. 153, no. 3, pp. 451-458, 2006.

[31] H. R. Karshenas, H. Daneshpajooh, A. Safaee, P. Jain, and A. Bakhshai, "Bidirectional dc-dc converters for energy storage systems," *Energy storage in the emerging era of smart grids,* vol. 18, 2011.

[32] L.-S. Yang and T.-J. Liang, "Analysis and implementation of a novel bidirectional DC–DC converter," *IEEE transactions on industrial electronics,* vol. 59, no. 1, pp. 422-434, 2011.

[33] F.-S. Pai and S.-J. Huang, "Design and operation of power converter for microturbine powered distributed generator with capacity expansion capability," *IEEE Transactions on Energy Conversion,* vol. 23, no. 1, pp. 110-118, 2008.

[34] N. M. Abdel-Rahim and J. E. Quaicoe, "Analysis and design of a multiple feedback loop control strategy for single-phase voltage-source UPS inverters," *IEEE Transactions on power electronics,* vol. 11, no. 4, pp. 532-541, 1996.